\documentclass{jaa}
\usepackage{natbib}
\usepackage{graphicx}

\newcommand{\ty}[1]{\mbox {TYC~8351-1081-1}}
\newcommand{\as}[1]{\mbox {ASAS~J210406-0522.3}}
\newcommand{\tyc}[1]{\mbox {TYC~8351-1377-1}}

\begin{document}\sloppy

\title{Analysis of Dual Band and Survey Photometry of Two Low Mass Ratio Contact Binary Systems}


\author{Surjit S. Wadhwa\textsuperscript{1*}, Ain Y. De Horta\textsuperscript{1}, Miroslav D. Filipovi\'c \textsuperscript{1} and Nick F. H.  Tothill\textsuperscript{1}}
\affilOne{\textsuperscript{1}School of Science, Western Sydney University, Locked Bag 1797, Penrith, NSW 2751, Australia.\\}


\twocolumn[{

\maketitle

\corres{19899347@student.westernsydney.edu.au}


\begin{abstract}
The study presents photometric analysis of the completely eclipsing contact binary systems \ty. and \as.. \ty. is an extremely low mass ratio (q=0.086) system with high degree of contact (f=0.66) while \as. is found to be in marginal contact (f=0.08) with a relatively low mass ratio of 0.272. There is good thermal contact in both systems with only a small difference in the temperature of the components. The systems have been observed by a number sky surveys over the past 20 years. We compare the light curve solutions from up to three of these surveys and find that survey photometric data manually analysed is robust and yields results comparable to dedicated ground based photometry. There is evidence of significant luminosity transfer from the primary to the secondary, on the order of $0.5L_{\bigodot}$ for \ty. and $0.06L_{\bigodot}$ for \as.. There appears to be no change in period of either system over the past 20 years and theoretical angular momentum loss is below current measurement threshold in both cases. We also show that the mass ratio and separation are well above the theoretical values for orbital instability in both cases. As would be expected, the density of the secondary components is significantly higher relative to the primary. 
\end{abstract}


}]


\doinum{12.3456/s78910-011-012-3}
\artcitid{\#\#\#\#}
\volnum{000}
\year{0000}
\pgrange{1--}
\setcounter{page}{1}
\lp{1}

\section{Introduction}
Contact binary systems are common with nearly 500000 listed in The International Variable Star Index (VSX) \citep{2006SASS...25...47W}. Most of them have only been relatively recently recognised through all sky surveys such as the Automated All Sky Survey (ASAS) \citep{2002AcA....52..397P}, All Sky Automated Survey - Super Nova (ASAS-SN) \citep{2014ApJ...788...48S, 2020MNRAS.491...13J}, Catalina Sky  Survey \citep{2017MNRAS.469.3688D} among many others. Since the recognition that luminous transient events dubbed red novae are the result of contact binary component mergers \citep{2011A&A...528A.114T} there has been increasing interest in their study not only from the point of view detecting pre-merger candidates \citep{2021MNRAS.501..229W} but also with a view to understanding evolutionary and astrophysical processes such as energy transfer and angular momentum loss \citep{2020MNRAS.493.1565D, 2016AJ....151...69S, 2004A&A...426.1001C}. 

Given the sheer sample size it is impractical to attempt to individually study a large sample of system in the hope of finding unstable systems or understand certain astrophysical principals as they may pertain to the contact binary systems as a whole. Recently there have been some forays into using the photometric data from sky surveys to analyse large number of systems through automated processes \citep{2020ApJS..247...50S}. There have been only a few investigations into the validity of using survey photometry as a satisfactory tool to obtain accurate light curve solutions. \citet{2020ApJS..247...50S} suggested that ASAS-SN photometric data offered good agreement for determining the mass ratio relative to ground based observations while \citet{2020MNRAS.493.1565D} demonstrated good agreement with respect to the mass ratio and fractional radii between dedicated and survey data with some variation in fill-out factor and the temperature ratio. Comparison of astrophysical processes such as energy transfer, luminosity ratio, orbital stability and effects of different photometry bands between dedicated ground based and survey photometry is lacking.

As part of a broader project dealing with orbital stability and identification of low mass ratio contact binary systems we are developing a semi-automated algorithm to select potential candidates from survey photometric data. Recently we identified two potential low mass ratio contact binary systems using the algorithm and undertook ground based observations. In this report we present the results along with comparisons with survey data and review of various astrophysical properties.

\ty. ($\alpha_{2000.0} = 17\ 41\ 00.25$, $\delta_{2000.0} = -48\ 51\ 45.9$) (T8351) was discovered as contact binary by ASAS with a period of 0.448058d and reported visual magnitude amplitude of 0.24mag. The system appears in the ASAS-SN survey as ASASSN-V J174100.31-485145.7 however is mis-classified as an RRC system with a period 0.2240307d and amplitude of 0.22mag. 

\as. (A2104) ($\alpha_{2000.0} = 21\ 04\ 05.56$, $\delta_{2000.0} = -05\ 22\ 21.5$) (A2104) was discovered as contact binary by ASAS with a period of 0.284460d and reported visual magnitude amplitude of 0.45mag. The system appears in the Catalina survey database as CRTS	J210405.5-052221 with a similar V band amplitude and period. Automated analysis of A2104 by \citet{2020ApJS..247...50S} indicates a mass ratio of 0.21 and a shallow fill-out of 11\%. 

We report the first ground based dual band (V and R) photometry and formal light curve analysis of both systems. In addition, we collate most available survey photometric data (different bands) from ASAS, Catalina Survey, ASAS-SN, Transiting Exoplanet Survey Satellite (TESS) Extended Mission \citep{2021RNAAS...5..234K} and the Zwicky Transient Facility (ZTF) \citep{2019PASP..131a8003M} and carry out light curve analysis on each available data set for both systems. We explore the orbital stability of the systems and examine some astrophysical properties such as the energy (luminosity) transfer, bolometric luminosity ratio, theoretical angular momentum loss estimation and density of the components. We compare and contrast the light curve solutions, absolute parameters and astrophysical indicators between the different data sets as an indicator of suitability of survey photometric data for analysing contact binary light curves.

\section{Photometry and Light Curve Analysis}
Ground based images using standard Johnson V and R band filters were acquired with the SBIG 8300T CCD camera attached to the Western Sydney University 0.6m telescope. All images were appropriately calibrated with flat, dark and bias frames and differential photometry performed using AstroImageJ \citep{2017AJ....153...77C}. Observation dates, number of observations, exposure times and details of comparison and check stars are summarised in Table 1. In addition, Table 1 also summarises duration of observations of survey data. Single time of minima were observed for each system and determined using the method of \citet{1956BAN....12..327K} at 2459432.08297$\pm 0.00594$ for T8351 and \mbox{HJD~2459442.08631 $\pm 0.00038$} for A2104. Orbital period for each system was revised using the Peranso software package \citep{2016AN....337..239P} to \mbox{0.448073d $\pm 0.000005$} for T8351 and \mbox{0.284457d $\pm 0.000003$} for A2104. The light curve was folded with the newly derived elements and normalised to the maximum brightness. Survey photometry data is available from ASAS, ASAS-SN and TESS for T8351 and ASAS-AS, Catalina and ZTF (r' and g' bands) for A2104. ASAS data for A2104 was too scattered and not suitable for light curve analysis but was suitable for period analysis (see below). 
 \begin{table*}
 \centering
	
	\label{tab:JAA T1}
	\begin{tabular}{|c|c|c|}
          \hline
          Parameter & T8351&A2104\\
          \hline
          Observation Period&29 July 2021 - 17 August 2021&12-17 August 2021\\
          Total Observations&685(V), 535(R)&429(V), 401(R)\\
          Exp Time&60s(V), 45s(R)&60s(v), 45s(R)\\
          Comp Star&2MASS-J17404638-4855016&2MASS-J2104086-0522159\\
          Chk Star&2MASS-J17411898-4846148&2MASS-J2104126-0521150\\
          ASAS (Obs Period)&Feb 2001 - Oct 2009&April 2001 - Sept 2009\\
          ASAS-SN (Obs Period)&March 2016 - Sept 2018&April 2014 - Oct 2018\\
          Catalina (Obs Period)&Not Observed&May 2005 - Oct 2013\\
          ZTF (Obs Period)&Not Observed&May 2018 - June 2021\\
          TESS (Obs Period)&May 2019 - Jun 2019&Not Observed\\
          
                      \hline
	\end{tabular}
	\caption{Observational parameters for T8351 and A2104}
	\end{table*}
The geometric parameters including the mass ratio, inclination and fill-out are highly correlated such that reliable photometric light curve solutions of contact binary systems is only achievable in the presence of  complete eclipses \citep{2005Ap&SS.296..221T}. The ground based (and survey photometry) for both systems show complete eclipses and accurate light curve solution is therefore likely. Light curve modelling of dedicated and survey photometry data was carried out using the 2009 version of the Wilson-Devinney (WD) code with Kurucz atmospheres as incorporated into the Windows front end utility WDwin56d \citep{1990ApJ...356..613W, 1998ApJ...508..308K, 2021NewA...8601565N}. Simultaneous V and R band solutions were obtained for all ground based photometry. There is no significant asymmetry in brightness at phases 0.25 and 0.75 as such only unspotted solutions were modelled. Given the common envelope configuration of contact binary systems certain parameters can be fixed during light curve modelling. The bolometric albedos $A_1 = A_2 = 0.5$ and the gravity darkening coefficients $g_1 = g_2 = 0.32$ were fixed \citep{1967ZA.....65...89L} and simple reflection treatment was applied \citep{1969PoAst..17..163R}. Logarithmic limb darkening coefficients \citep{2015IBVS.6134....1N} were interpolated from \citet{1993AJ....106.2096V}. We note that the TESS imaging bandpass is quite broad 600 - 1000nm centered on the traditional Cousins I band (786.5nm) \citep{2015JATIS...1a4003R}. We modelled the TESS photometry using infrared limb darkening coefficients. Based on the J-H values reported in the VSX database along with the calibration for low mass (F3-K9) main sequence stars of \citet{2013ApJS..208....9P} we interpolated the mass and effective temperature of the primaries ($M_1 , T_1$) as $1.47M_{\bigodot}$, 6720K for T8351 and $0.77M_{\bigodot}$, 5090K for A2104. The adopted values are within 150K based on distance estimation of 452.9Pc (T8351) and 291.0Pc (A2104) \citep{2018yCat.1345....0G}.

The well-established mass ratio search grid method \citep{1982A&A...107..197R} was used in modelling all light curves. Inclination ($i$), potential ($\Omega$), temperature of the secondary ($T_2$) and luminosity of the primary ($L_1$) were the adjustable parameters for several fixed values of the mass ratio (q) from 0.06 to 1.0. The mass ratio search grids are illustrated in panel B of Figures 1 and 2 (only the portion near the minimum error is illustrated for clarity). Similar process was applied to all survey photometric data for both systems. In the final iteration the mass ratio was also made a free parameter and the standard deviations from the resulting output were recorded as the errors for each parameter. All subsequent error estimations were propagated from these values. We used the square of the sum of the residuals (as reported by the WD code) to calculate the residual standard deviation for each solution as a measure of the quality of fit between the various data sources. As expected the the standard deviation of the residual is smaller for dedicated observations compared to survey data except in the case of TESS data which showed minimal scatter and provided the best fit for T8351.  The light curve solutions are summarised in Tables 2 and 3. Observed and fitted light curves are illustrated in panel A and three dimensional representation in panel C of Figures 1 and 2. 

\begin{figure*}[!ht]
    \label{fig:JAAFIG1}
	\includegraphics[width=\textwidth]{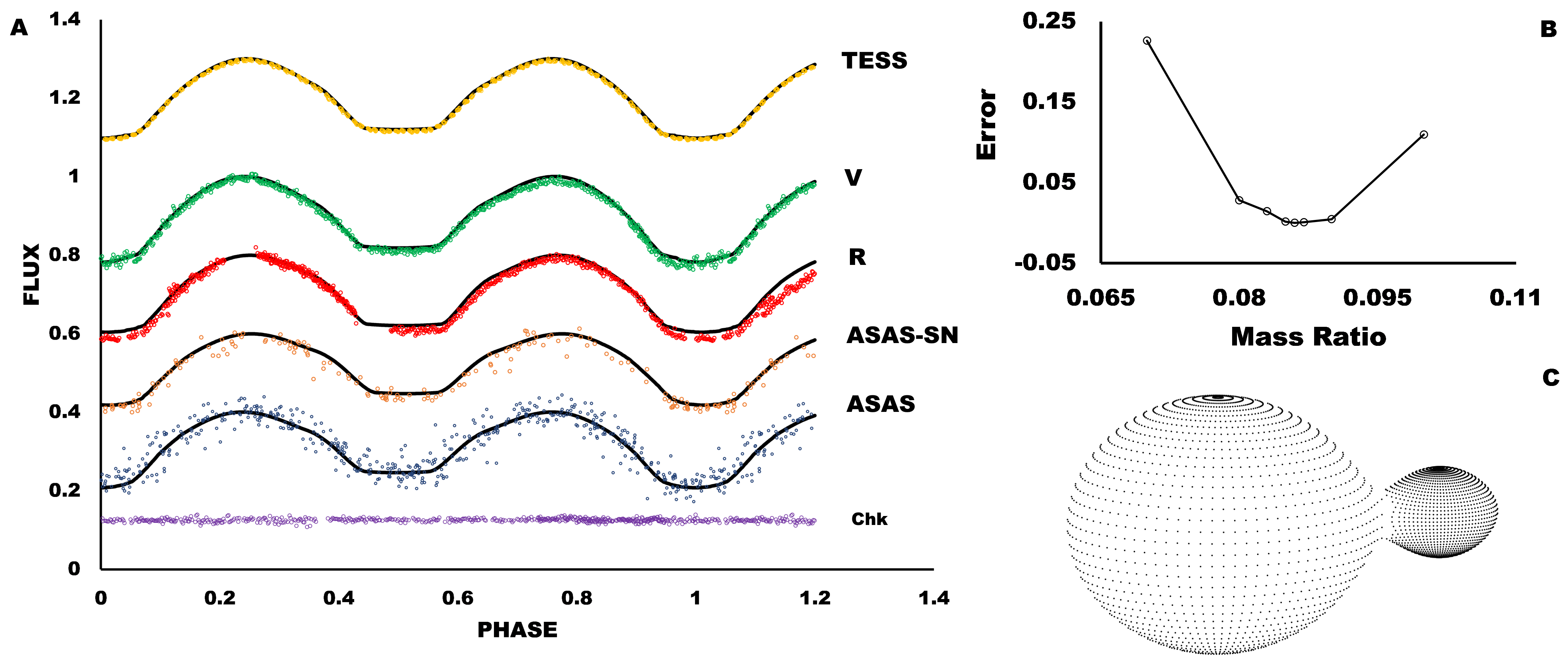}
    \caption{The left panel (A) illustrates the observed and fitted light curves for the simultaneous solution for V and R bands along with that of the check star (marked "chk"). Also plotted are the survey photometry and fitted curves for ASAS-SN, ASAS and TESS. All observed curves were normalised to the maximum brightness. The plots have been moved randomly in the vertical axis for clarity. The top half of the right panel (B) illustrates the mass ratio search grid for T8351 for the simultaneous V and R band solution. The error has been normalised to the lowest error value. The bottom of the right side (C) is a three dimensional representation of T8351 based on the current study parameters.}
    \end{figure*}

\begin{figure*}[!ht]
    \label{fig:JAAFIG2}
	\includegraphics[width=\textwidth]{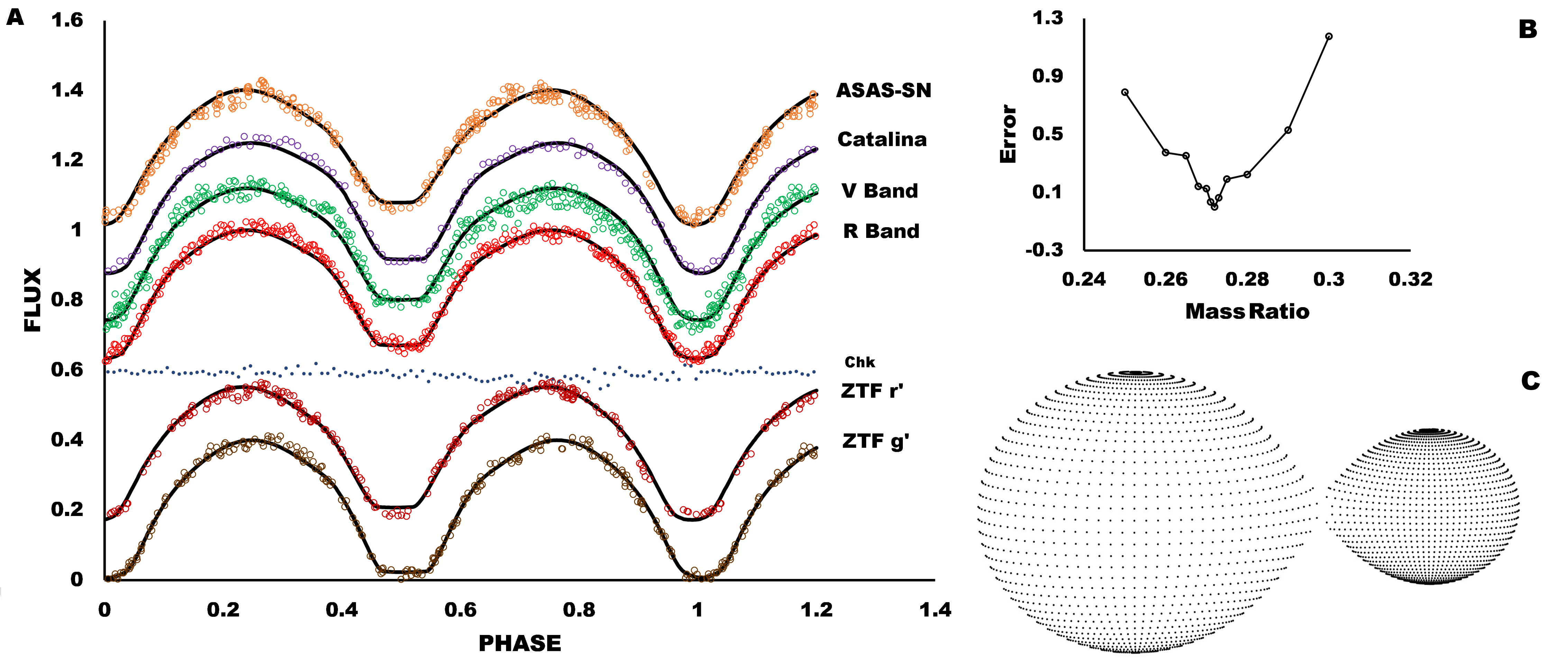}
    \caption{The left panel (A) illustrates the observed and fitted light curves for the current study (middle graphs) along with that of the check star (marked "chk"). Also plotted are the survey photometry and fitted curves for ASAS-SN, Catalina, ZTF r'and ZTF g'. All observed curves were normalised to the maximum brightness. The plots have been moved randomly in the vertical axis for clarity. The top half of the right panel (B) illustrates the mass ratio search grid for the simultaneous V and R band solution. The error has been normalised to the lowest error value. The bottom of the right side (C) is a three dimensional representation of A2104 based on the current study parameters.}
    \end{figure*}

The major features of note are strong agreement in the solutions between dedicated and survey photometry particularly with respect to the mass ratio. The maximum variation in the mass ratio between survey and dedicated photometry is 0.006 for T8351 and 0.013 for A2104. Interestingly the Catalina Survey data (A2104) solved manually yields a mass ratio of 0.27 as opposed the the automated solution \citep{2020ApJS..247...50S} with a mass ratio 0.21. All models suggest a cooler secondary for both systems except the ZTE g'data which indicates the secondary marginally hotter in the case of A2104. As can be seen from Tables 2 and 3 there was good agreement with respect other physical parameters such as fractional radii and inclination between dedicated and survey photometry for both systems. Additionally, we confirm the finding of \citet{2020MNRAS.493.1565D} that some variation in fill-out is likely between survey and dedicated photometry. This is possibly due to the scatter associated with survey data as the shape of the shoulders near the maxima may be influenced by the fill-out fraction \citep{1993PASP..105.1433R}.

\section{Absolute Parameters and Astrophysical Considerations}
\subsection{Absolute Parameters}
It is well accepted that the primary components of contact binary systems resemble zero age main sequence stars (ZAMS) \citep{2013MNRAS.430.2029Y}. As noted above we used a recent ZAMS calibration to estimate $M_1$ and $T_1$. Mass of the secondary ($M_2$) can be derived from the mass ratio as 0.13$M_{\bigodot}$ (T8351) and 0.21$M_{\bigodot}$ (A2104). As both components in a contact binary system overflow their Roche lobes their relative radii are not independent but determined by the Roche geometry and mass ratio. The light curve solution provides fractional radii of the components ($r_{1,2}$) in three orientations. We used the geometric mean of these to estimate both $r_1$ and $r_2$. Using Kepler's third law we determined the separation ($A$) in solar units and from the separation the absolute radii of the components as per \citep{2005JKAS...38...43A} $R_{1,2}$ = $A\times r_{1,2}$. The light curve solution also provides the luminosity of each component as a dimensionless number from which we calculate the ratio of $L_2/L_1$ as 0.107 for T8351 and 0.2945 for A2104 for current observed light curve solutions. The luminosity ratios for survey photometry along with the absolute parameters of the system are summarised in Tables 2 and 3. 

    \subsection{Mass Transfer, Angular Momentum Loss and Orbital Stability}
The mainstay of conservative mass transfer is a change in period. A decreasing period indicates transfer of mass from the primary to the secondary and vice versa for an increasing period. Period variations are usually small and require dedicated (high cadence) long term (decades) observations. Given the spread of the survey observational data and poor cadence a traditional period analysis of either system is not feasible. T8351 was observed over 6700 days by the ASAS, ASAS-SN and TESS surveys while A2104 was observed by ASAS, ASAS-SN, Catalina and ZTF for over 7000 days. As reported by \citet{1996ApJ...460L.107S}, period analysis of uneven observational data can be successfully achieved by employing periodic orthogonal polynomials to fit observations along with an analysis of variance statistic as a marker of the quality of the fit. The method was used by \citet{2011A&A...528A.114T} to determine the change in period of V1309 Sco from scattered survey photometric data spanning a number of years. They divided the entire Optical Gravitational Lensing Experiment (OGLE) data-set spanning over 8 years into multiple subsets and determined periods for each to show that there had been a significant decrease in the period of V1309 Sco prior to its merger event. We use the same method to explore period variations in the survey photometric data. In addition, we checked for potential alias periods by performing spectral window and false alarm analysis as described in the Peranso software user guide. The full photometry data set of all surveys was divided into multiple overlapping subsets with an average duration of 616.7 days (range: 15 - 909) and an average of 193 observations (range: 102 - 536) per subset for T8351 and 533 days (range: 285 - 957) and 142 observations (range: 77 - 212) per subset for A2104. Period search was carried out for each subset and plotted against the mid subset date. For all subsets the period selected was the one with the highest degree of fit with no detectable alias periods. The maximum reported potential false alarm for any one subset was less than 0.5\%. There appears to have been no significant change in the period for either system with the maximum variation of 0.00002d for T8351 and 0.000006d for A2104. The period trends are is illustrated in Figure 3. Lack of significant change in period goes against mass transfer suggesting likely stable configuration. The stability of the period however does not explain any period decrease due to angular momentum loss (AML). We discuss this next.

Contact binary systems typically have periods between 0.2 and 1 day with synchronised rotations. As rotation rate increases magnetic activity also increases \citep{2019BlgAJ..31...97G}. The effect of the increased magnetic field is to decrease the spin through torquing of the system by mass overflow. Given the rapid spin, the magnetized stellar wind becomes twisted resulting in the trapping and dragging of charged particles along the magnetic field lines. This results in the transfer of angular momentum from the system to the charged particles which in turn leads to a reduction in the spin and period of the system.  \citet{1994ASPC...56..228B} deduced a theoretical constraint on the rate of AML as per the following equation:

\begin{equation}
\begin{split}
 \frac{dP}{dt} \approx 1.1\times 10^{-8}q^{-1}(1+q)^2\times (M_1 + M_2)^{-5/3}k^2\times\\
 (M_1R_1^4 + M_2R_2^4)P^{-7/3}
 \end{split}
\end{equation}
where q is the mass ratio ($M_2/M_1$), $M_{1,2}$ are masses of the primary and secondary in solar units, $R_{1,2}$ are the radii of the primary and secondary components, $P$ is the period in days and $k$ is the gyration radius of the primary. The resultant rate change in the period is in days  yr$^{-1}$.

By interpolating the value of $k^2$  from \citet{2009A&A...494..209L} for a rotating and tidally distorted stars of 1.47$M_{\bigodot}$ and 0.77$M_{\bigodot}$ we calculated the current expected rate of change in the period due AML for T8351 as $-1.9\times10^{-7}$ days yr$^{-1}$ and for A2104 as $-7.9\times10^{-8}$ days yr$^{-1}$. Assuming a constant decrease at this rate, this equates to a total decrease in the period of 0.28 seconds over the entire observation period for T8351 and 0.14 seconds for A2104. Such a small total decline in the period is well beyond the resolution of the available data so it is not surprising that no significant change in period has been detected. In the case of T8351 there is a suggestion of slight increase in the period indicating other factors such as ongoing mass transfer, apsidal motion, light travel time effects or magnetic activity cycles maybe the main drivers of period variation at this time \citep{2018MNRAS.474.5199L}.

The minimal expected AML suggests a likely stable configuration. \citet{2021MNRAS.501..229W} introduced a set formulae linking the orbital stability of contact binary systems with the mass ratio and separation. They showed that the mass ratio and separation at which orbital instability was likely is dependent on the mass of the primary for low mass contact binary systems. Their modelling yielded simple quadratic relationships linking the mass of the primary and the instability mass ratio for fill-out of 0 and 1.0. These two relations thus provide a range of mass ratios over which instability is possible. Systems with mass ratios above this range are highly unlikely to exhibit orbital instability. Secondarily they derived a numerical relationship linking the mass ratio, radius of the primary and the fill-out to determine the instability separation. Using these we determine the instability mass ratio for T8351 to be between 0.049 and 0.054 and for A2104 to be between 0.16 and 0.20 compared to the current values of 0.086 and 0.272 respectively. The instability separations based on current observed light curve solutions are determined to be 2.23$R_{\bigodot}$ and 1.63$R_{\bigodot}$ respectively both significantly smaller than the current separations of 2.87$R_{\bigodot}$ and 1.81$R_{\bigodot}$ respectively. The mass ratios are higher and separations wider than theoretical values at which orbital instability is likely, we therefore conclude that the systems are currently stable and are likely to remain that way on a nuclear time scale.

\subsection{Luminosity Transfer and Component Densities}

Using the calibration of \citet{2013ApJS..208....9P} we estimated the absolute luminosity of the primary component and from the luminosity ratio the absolute luminosity of the secondary. The estimated luminosities of the secondaries are considerably brighter than their ZAMS counterparts. This is a usual finding in contact binary systems and is due to the transfer of luminosity (energy) from the primary to the secondary \citep{1968ApJ...151.1123L}. While studying the luminosity transfer in contact binaries  \citet{2004A&A...426.1001C} found that the energy transfer rate was a function of mass and bolometric luminosity ratio and introduced the energy transfer parameter defined as:
\begin{equation}
    \beta = \frac{L_{1,Obs}}{L_{1,ZAMS}}
\end{equation}
In addition they showed that there exists a minimum transfer parameter for contact binaries defined as:
\begin{equation}
    \beta_{min} = \frac{1}{1+\Big(\frac{L_2}{L_1}\Big)_{bol}}.
\end{equation}
where ($L_2/L_1$)$_{bol}$ is the bolometric luminosity ratio defined as:
\begin{equation}
    \bigg(\frac{L_2}{L_1}\bigg)_{bol} = \frac{L_2}{L_1}\times 10^{0.4\times(BC_1 - BC_2)}
\end{equation}
where $BC_{1,2}$ are bolometric corrections. They concluded that for low mass ratio contact binary systems regardless of temperatures of the secondary the luminosity transfer was on or near the minimum transfer envelope. As noted by \citet{2004A&A...426.1001C} from the definition of the transfer parameter the amount of luminosity transferred from the primary to the secondary can be expressed as $\Delta L = (1-\beta)L_1$. 

As noted by \citet{2003ASPC..293...76W} luminosity transfer from the primary to the secondary will restructure the secondary to make it oversized and over-luminous for its mass. Similarly the density and effective temperature of the secondary are also affected by the luminosity transfer. We interpolated the bolometric corrections for low mass stars from \citet{2013ApJS..208....9P} and calculated the transfer parameter for each system. We found that both systems lie close to or on the minimum luminosity transfer envelope. Based on the calibration of \citet{2013ApJS..208....9P} the solar luminosities for T8351 and A2104 are $5.12L_{\bigodot}$ and $0.28L_{\bigodot}$ respectively. The amount of transferred luminosity from the primary to the secondary is considerable at approximately $0.49L_{\bigodot}$ for T8351 and $0.064L_{\bigodot}$ for A2104. The transfer parameter was plotted against the bolometric luminosity ratio and is shown in Figure 4 while the luminosity transfers are summarised in Tables 2 and 3. We have not provided error estimations for the luminosity transfer as we have used calibrated or fixed values for all relevant variables.

\begin{figure*}
    \label{fig:JAAFIG3}
	\includegraphics[width=\textwidth]{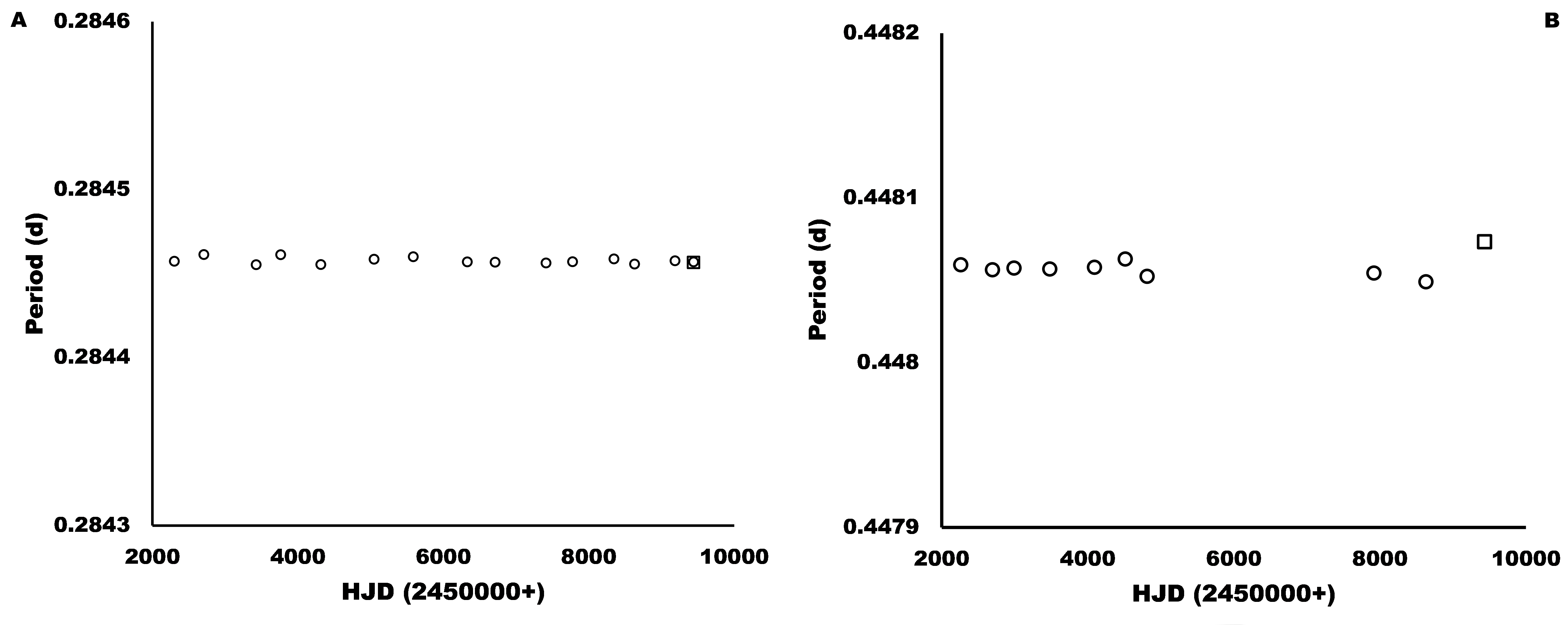}
    \caption{Both systems have a relatively stable period over the past 20 years of survey observations. The period for A2104 is illustrated on the left (Panel A) and for T8351 on the right (Panel B). The square box in each panel represents the period from current observations}
    \end{figure*}
    
    \begin{figure}
    \label{fig:JAAFIG4}
	\includegraphics[width=\columnwidth]{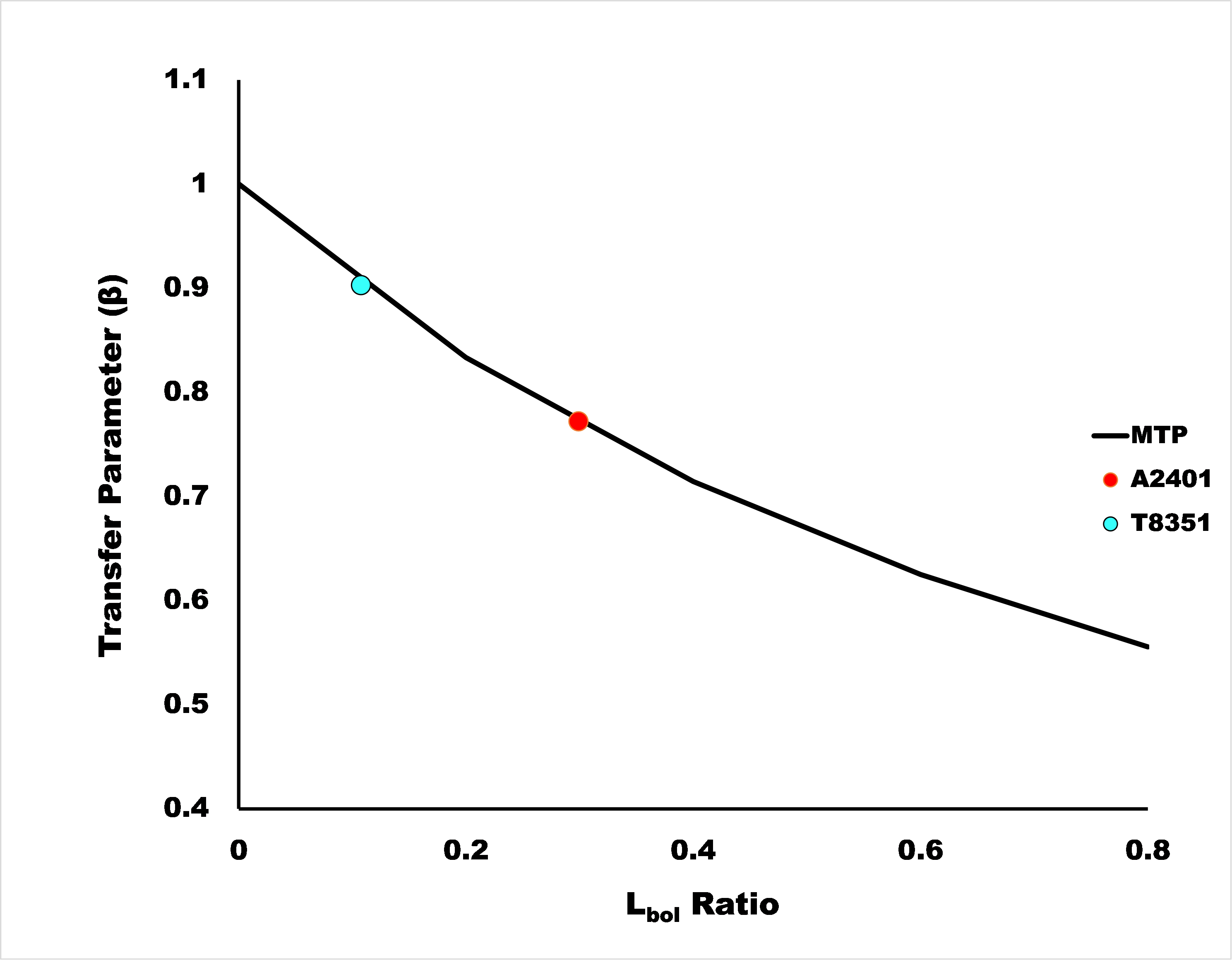}
    \caption{Illustration of the bolometric luminosity ratio against the transfer parameter. The solid line represents the minimum transfer parameter (MTP). Both systems being of relatively low mass ratio are expected lie close to the minimum transfer parameter}
    \end{figure}

We confirm the findings of \citet{2013MNRAS.430.2029Y} that radius of the primary is larger than a corresponding ZAMS and that the luminosity and radius of the secondary are considerably larger than the corresponding ZAMS of same mass. Clearly the secondaries are quite distorted and as argued by \citet{2013MNRAS.430.2029Y} they must have had a different evolutionary path relative to ZAMS. It is postulated \citep{2013MNRAS.430.2029Y} that during the formation and evolution of contact binary systems the current primary was initially a lower mass secondary which gained mass from the initial primary as it evolved and expanded beyond its Roche lobe. As mass transfer continued the initial primary became smaller and the initial secondary larger such that the roles reversed and both stars grew beyond their Roche lobes to share a common envelope. The result of such a mass transfer is that the secondary retains higher mass core constituents and the current primary gains lower mass outer layer constituents. \citet{2004A&A...414..317K} argues that such a scenario will lead to significant variation in the density of the components with the secondary being always denser.

\citet{1981ApJ...245..650M} showed that the density (gcm$^{-2}$) of contact binary components can be expressed as a function of the period, radii and the mass ratio. The difference between the densities of the components can be expressed as follows:
\begin{equation}
    \Delta\rho = \frac {0.0189q}{R_2^3(1+q)P^2} - \frac {0.0189}{R_1^3(1+q)P^2}
\end{equation}
where $P$ is the period in days, $R_{1,2}$ radii of the components and $q$ is the mass ratio.
We calculated the difference in densities of the components for all light curve solutions, summarised in Table 2 and 3, and as expected the density of the secondary is significantly higher and no significant difference between light curve solutions.

 \begin{table*}
 \centering
	
	\label{tab:JAA T2}
	\begin{tabular}{|c|c|c|c|c|}
          \hline
          Parameter & Current&ASAS&ASAS-SN&TESS\\
          \hline
          $T_1$(K)(Fixed)&6720&6720&6720&6720\\
          $T_2$(K)&6480$\pm 14$&6270$\pm 65$&6390$\pm 26$&6520$\pm 10$\\
          q ($M_2/M_1$)&0.086$\pm 0.002$&0.085$\pm 0.006$&0.08$\pm 0.005$&0.084$\pm 0.001$\\
          Inclination($^{\circ}$)&80.15$\pm 0.5$&76.80$\pm 2.0$&75.40$\pm 1.9$&78.15$\pm 0.14$\\
          Fill-out($f$)&0.66$\pm0.04$&0.22$\pm 0.04$&0.16$\pm 0.05$&0.81$\pm 0.02$\\
          $r_1$(mean)&0.611&0.600&0.603&0.617\\
          $r_2$(mean)&0.220&0.203&0.197&0.227\\
          $M_1/M_{\bigodot}$(fixed)&1.47&1.47&1.47&1.47\\
          $M_2/M_{\bigodot}$&0.130$\pm 0.005$&0.12 $\pm 0.01$&0.12 $\pm 0.01$&0.12$\pm 0.01$\\
          $A$($R_{\bigodot}$)&2.87$\pm 0.02$&2.87$\pm 0.03$&2.86$\pm 0.03$&2.87$\pm 0.02$\\
          $R_1/R_{\bigodot}$&1.76$\pm 0.02$&1.72$\pm0.03$&1.73$\pm0.03$&1.77$\pm 0.02$\\
          $R_2/R_{\bigodot}$&0.63$\pm 0.02$&0.58$\pm 0.03$&0.56$\pm 0.03$&0.65$\pm 0.02$\\
          $L_2/L_1$&0.107&0.085&0.087&0.110\\
          Luminosity Transfer ($L_{\bigodot}$)&0.49&0.42&0.42&0.52\\
          $\Delta \rho$&-0.32&-0.49&-0.51&-0.25\\
          Std Dev (Res)&0.005 (V), 0.008 (R)&0.02&0.01&0.002\\
                      \hline
	\end{tabular}
	\caption{Light curve solution, absolute parameters, luminosity transfer and density difference for T8351. Results are for the current study, ASAS, ASAS-SN and TESS Surveys.}
	\end{table*}

 \begin{table*}
 \centering
	
	\label{tab:JAA T3}
	\begin{tabular}{|c|c|c|c|c|c|}
          \hline
          Parameter & Current&Catalina&ASAS-SN&ZTF r'&ZTF g'\\
          \hline
          $T_1$(K)(Fixed)&5090&5090&5090&5090&5090\\
          $T_2$(K)&5035$\pm 18$&5065$\pm 24$&4915$\pm 26$&5050$\pm 27$&5220$\pm 16$\\
          q ($M_2/M_1$)&0.272$\pm 0.003$&0.270$\pm 0.006$&0.268$\pm 0.006$&0.281$\pm 0.007$&0.278$\pm 0.003$\\
          Inclination($^{\circ}$)&84.5$\pm 1.1$&83.2$\pm 1.9$&80.3$\pm 1.2$&83.4$\pm 1.3$&89.8$\pm 3.2$\\
          Fill-out($f$)&0.08$\pm 0.02$&0.01$\pm 0.02$&0.20$\pm 0.05$&0.19$\pm 0.05$&0.16$\pm 0.04$\\
          $r_1$(mean)&0.500&0.497&0.507&0.502&0.502\\
          $r_2$(mean)&0.276&0.271&0.282&0.285&0.283\\
          $M_1/M_{\bigodot}$(fixed)&0.77&0.77&0.77&0.77&0.77\\
          $M_2/M_{\bigodot}$&0.209$\pm 0.004$&0.208 $\pm 0.01$&0.206 $\pm 0.01$&0.216 $\pm 0.011$&0.214 $\pm 0.01$\\
          $A$($R_{\bigodot}$)&1.81$\pm 0.01$&1.81$\pm 0.01$&1.80$\pm 0.01$&1.81$\pm 0.01$&1.81$\pm 0.01$\\
          $R_1/R_{\bigodot}$&0.90$\pm 0.01$&0.90$\pm 0.01$&0.91$\pm 0.01$&0.91$\pm 0.01$&0.91$\pm 0.01$\\
          $R_2/R_{\bigodot}$&0.49$\pm 0.01$&0.49$\pm 0.01$&0.51$\pm 0.01$&0.52$\pm 0.01$&0.51$\pm 0.01$\\
          $L_2/L_1$&0.294&0.293&0.258&0.310&0.360\\
          Luminosity Transfer ($L_{\bigodot}$)&0.056&0.056&0.052&0.058&0.063\\
          $\Delta \rho$&-0.91&-0.99&-0.79&-0.77&-0.8\\
          Std Dev (Res)&0.01 (V), 0.01 (R)&0.01&0.03&0.02&0.01\\
                      \hline
	\end{tabular}
	\caption{Light curve solution, absolute parameters, luminosity transfer and density difference for A2104. Results are for the current study, Catalina Survey, ASAS-SN Survey and the ZTF survey (r'and g' bands)}
	\end{table*}

	\section{Summary and Conclusions}
Contact binary systems are common and have attracted greater interest over the last 30 years as they present an ideal astrophysical laboratory to explore stellar evolution, orbital stability and astrophysical processes such as energy transfer and AML. As part of an ongoing project we aim to identify potential low mass ratio contact binary systems with a view to performing light curve analysis and studying astrophysical properties leading to a better understanding of their structure, stability and evolutionary outcomes. The present study presents the photometric analysis of T8351 and A2104, both of which have been recently recognised as contact binary systems and identified by us as potential low mass ratio candidates. We confirm that both systems have relatively low mass ratios with good thermal contact between the components. T8351 displays relatively deep contact (66\%) while A2106 is a marginal contact system. Both systems demonstrate significant energy (luminosity) transfer from the primary to the secondary. Evidence for significant angular momentum loss is lacking with no appreciable change in the period during the period of observations spanning over 20 years. The mass ratio and separation for both systems are well above theoretical levels for orbital instability and as such we consider both systems to be relatively stable and not potential merger candidates. Like most contact binary systems the secondary component has significantly greater luminosity, size and density relative to main sequence stars of similar mass. 

Both systems have been observed by multiple surveys over the past 20 years and manual analysis of survey photometry presented in this study yields light curve solutions similar to dedicated observation data however differ somewhat from automated survey data analysis in the case A2106. Concordance of dedicated and survey photometric analysis confirms similar findings by \citet{2020MNRAS.493.1565D} and \citet{2021AJ....162...13L} and suggests that survey photometry offers a resource that can be successfully employed in the identification of potentially interesting binary systems. We hope to utilise survey photometry as a vehicle in the identification of potential merger candidates in future studies targeting contact binary systems.

\section*{Acknowledgements}

Based on data acquired on the Western Sydney University, Penrith Observatory Telescope. We acknowledge the traditional custodians of the land on which the Observatory stands, the Dharug people, and pay our respects to elders past and present.\\

This research has made use of the SIMBAD database, operated at CDS, Strasbourg, France.
\vspace{-1em}


\bibliography{P1.bib}



\end{document}